\documentclass[10pt]{article}

\usepackage{a4wide}
\usepackage{authblk}
\usepackage{graphicx}
\usepackage{epstopdf, epsfig}	
\usepackage{amsmath}
\usepackage{amssymb}
\usepackage[round,sort&compress,comma,authoryear]{natbib}
\usepackage{float}
\usepackage{hyperref}
\usepackage{lscape}
\usepackage{newtxtext}
\usepackage{newtxmath}
\usepackage{natbib}
\usepackage{hyperref}
\hypersetup{
    colorlinks = true,
    urlcolor   = blue,
    citecolor  = black,
}

\newcommand{\RomanNumeralCaps}[1]
\linenumbers

\newcommand\Mor{\mbox{\textit{Mo}}}  % Morton number
\newcommand\Web{\mbox{\textit{We}}}  % Weber number
\usepackage[svgnames]{xcolor} % De la couleur
\definecolor{eau}{rgb}{1,     0,     0}
\definecolor{ethanol}{rgb}{0.3, 0, 0.5}
\definecolor{eau_gluc_1}{rgb}{0.5, 0.5, 0.5}
\definecolor{eau_gluc_2}{rgb}{0.35, 0.35, 0.35}
\definecolor{eau_gluc_3}{rgb}{0, 0, 0}
\definecolor{eau_ucon_8}{rgb}{0.2422,    0.1504,    0.6603}
\definecolor{eau_ucon_17}{rgb}{0.2802,    0.2764,    0.9204}
\definecolor{eau_ucon_42}{rgb}{0.2791,    0.3475,    0.9733}
\definecolor{eau_ucon_80}{rgb}{0.2473,    0.4319,    0.9983}
\definecolor{eau_ucon_140}{rgb}{ 0.1573 ,   0.5854,    0.9254}
\definecolor{eau_ucon_152}{rgb}{0.0433,    0.7039,    0.8242}
\definecolor{eau_ucon_210}{rgb}{0.1853,    0.7721 ,   0.6379}
\definecolor{eau_ucon_260}{rgb}{0.3548,    0.8016,    0.4669}
\definecolor{eau_ucon_620}{rgb}{0.8389 ,   0.7457,    0.1546}
\definecolor{eau_ucon_930}{rgb}{0.9906,    0.8095,    0.1906}
\definecolor{eau_ucon_1120}{rgb}{0.9608,    0.8902,    0.1532}
\definecolor{eau_ucon_2890}{rgb}{0.9769,    0.9839,    0.0805}
\definecolor{Triton}{rgb}{0.7500,    0,    0.7500}
\definecolor{Filella}{rgb}{0.466, 0.674, 0.188}
\definecolor{Pavlov}{rgb}{0.929, 0.694, 0.125}

\newcommand\DEL[1]{\sout{}}

\title{Bubble rise in a Hele-Shaw cell: bridging the gap between viscous and inertial regimes}

\author[1]{Benjamin Monnet}
\author[1]{Christopher Madec}
\author[1]{Val\'{e}rie Vidal} 
\author[1, 3, $\dagger$]{Sylvain Joubaud}
\author[2]{{J. John Soundar Jerome}}
\affil[1]{Univ Lyon, ENS de Lyon, CNRS, Laboratoire de Physique, 69367 Lyon Cedex 07, France}
\affil[2]{Univ Lyon, Univ Claude Bernard Lyon 1, CNRS, Ecole Centrale de Lyon, INSA Lyon, LMFA, UMR5509, 69622 Villeurbanne France}
\affil[3]{Institut Universitaire de France (IUF), 1 rue Descartes, 75005 Paris, France}
\affil[$\dagger$]{email : sylvain.joubaud@ens-lyon.fr}

\date{}

\begin{document}
\maketitle

\begin{abstract}
The rise of a single bubble confined between two vertical plates is investigated over a wide range of Reynolds numbers. In particular, we focus on the evolution of the bubble speed, aspect ratio and drag coefficient during the transition from the viscous to the inertial regime. For sufficiently large bubbles, a simple model based on power balance captures the transition for the bubble velocity and matches all the experimental data despite strong time variations of bubble aspect ratio at large Reynolds numbers. Surprisingly, bubbles in the viscous regime systematically exhibit an ellipse elongated along its direction of motion while bubbles in the inertia-dominated regime are always flattened perpendicularly to it.

\end{abstract}

\section{Introduction} \label{sec:intro}
Main characteristics such as the rise speed $v_b$ and shape for isolated bubbles in a liquid of infinite extent are now well established~\citep{Davies_1950,Harper_1972, Maxworthy_1996, Clift_2005, Tripathi_2015}. In a confined environment, the question of the transition between the viscous and the inertial regime is still open and will be tackled in this paper. Conventionally, such cases are studied by considering bubbles in a liquid contained in a Hele-Shaw cell, consisting in two plates separated by a very small distance, $h$ \citep{Maxworthy_1986, Filella_2015, Gaillard_2021} or in cylindrical tubes \citep{Danov_2021}.
In the case of a bubble confined in a Hele-Shaw cell, the appropriate Reynolds number that compares the inertial and viscous effects is $\mbox{Re}_{2h}=(\rho v_b d_2/\eta) (h/d_2)^2$ \citep{Batchelor_2000} where $\rho$ is the liquid density, $\eta$ the liquid dynamical viscosity and $d_2=2(A/\pi)^{1/2}$ the equivalent bubble diameter computed from $A$, the area occupied by the bubble in the plane of the plates. 

As first demonstrated by~\cite{Taylor_1959} and later by~\cite{Eck_1978}, \cite{Maxworthy_1986} and \cite{Tanveer_1987}, when surface tension effects are neglected, large elliptical bubbles ($d_2 \gg h$) in the viscous regime ($\mbox{Re}_{2h}\ll 1$) rise at the characteristic speed
\begin{equation} \label{eq:velocity_maxworthy}
    v_M=v_M^{\star}\frac{a}{b}, ~\text{with}~v_M^{\star}=\frac{\Delta \rho g h^2}{12 \eta},
\end{equation}
where $\Delta \rho=\rho - \rho_g$ with $\rho_g$ the gas density, $g$ the gravity, $a$ and $b$ respectively the bubble length in the direction (longitudinal) and perpendicular (transverse) to its movement. In the limit of an almost horizontal cell ($g$ here is the effective gravity), \cite{Eck_1978} and \cite{Maxworthy_1986} observed that $a \simeq b$ and $v_b \simeq v_M^{\star}$ .
By including surface tension effects, \cite{Tanveer_1987} demonstrated that the analysis of \cite{Taylor_1959} should lead to a wide variety of solutions for the bubble shape which cannot be simply determined. 
Much later, using the generalised Onsager's principle \citep{Doi_2011} along with the \cite{Park_1984} boundary condition at an elliptical bubble perimeter, \cite{Xu_2020} emphasised that a single rising bubble is either circular ($a = b$), or flattened ($a < b$) due to viscous Bretherton dissipation at the moving bubble boundary, {which accounts for the energy loss in the lubrication films between the bubble and the walls}. While this result is in accordance with the experimental observations of \cite{Eck_1978} for an inclined Hele-Shaw cell, all bubbles in \cite{Maxworthy_1986} are elongated in the longitudinal direction ($a > b$). The latter result is also seen in more recent investigations conducted by \cite{Madec_2020} in a vertical Hele-Shaw cell. Therefore, the parameters which govern the bubble shape, in particular, its aspect ratio and more generally, its intricate relation to its proper rise speed (Eq.~\ref{eq:velocity_maxworthy}) is still not well-established.

In the inertial regime ($\mbox{Re}_{2h} > 1$), \cite{Roig_2012} found experimentally that, in distilled water, the time-averaged bubble speed $v_b$ follows ${v_b \simeq v_i = }~\xi \sqrt{gd_2}$ where $\xi$ was later found to depend on the cell gap \citep{Filella_2015} {so that}
\begin{equation} \label{eq:inertial_speed}
    {v_i}=\alpha (3/2)^{1/6} (h/d_2)^{1/6}\sqrt{gd_2}= \alpha \sqrt{gd_3}.
\end{equation}
{Here,} $d_3=(6V/\pi)^{1/3}$ is the diameter of an equivalent spherical bubble of same volume $V$ and {the empirical coefficient} $\alpha~{ = 0.71}$ is now independent of $h$ {\citep{Filella_2015, Pavlov_2021}}. {While this result is analogous to isolated spherical cap bubbles in an unbounded liquid \citep{Davies_1950}, the physical origin of the constant $\alpha=0.71$ is still not clear.} {By studying} air bubbles in various glycerol/water solutions and colloidal suspensions of fine silica particles in water , respectively, \cite{Hashida_2019, Hashida_2020} proposed that the factor $\alpha$ should depend on liquid characteristics such as $\rho$, $\eta$ and $\gamma$ (its surface tension). Regarding the bubble shape, \cite{Roig_2012} and \cite{Filella_2015} observed that all bubbles are flattened with respect to the longitudinal direction ($\chi = a/b < 1$). Note that those results concern the time-averaged properties of the bubble since for high enough Reynolds numbers, the bubble wake undergoes destabilisation whereby path instabilities and shape oscillations occur during bubble rise in a Hele-Shaw cell \citep{Kelley_1997, Roig_2012, Wang_2014}.

In this context, we focus on the time-averaged bubble rise speed and shape in a vertical Hele-Shaw cell of different gaps for a wide variety of liquids (see Table~\ref{sol}). %The main goal of our work is to provide an unified point of view for the viscous to inertial transition of large Hele-Shaw bubbles.
First, we quantify the bubble speed transition from the viscous (Eq.~\ref{eq:velocity_maxworthy}) to the inertial (Eq.~\ref{eq:inertial_speed}) limit by systematically varying the bubble Reynolds number $\mbox{Re}_{2h}$ from 10$^{-4}$ to 10$^2$. Then, we investigate how the bubble shape changes from an ellongated ellipse ($a>b$) to a flattened bubble ($a<b$). Finally, we provide a scaling law for the bubble drag coefficient in both viscous and inertial regimes.

\begin{table}[ht!]
  \begin{center}
  \begin{tabular}{cccccccc} 
      Solution    &  Viscosity   & Density   &  Surface Tension  & gap  & Symbol & Morton number &  Time ratio \\[3pt]
       & $\eta$ (mPa.s) & $\rho$ (kg.m$^{-3}$) & $\gamma$ (mN.m$^{-1}$) & $h$ (mm) &  & $\Mor=g\eta^4/(\rho \gamma^3)$  & $\tau/T$ \\
       & & & & & & & \\
      water    & 0.94 & 997 & 72 &2.3 & \color{eau} \large{$\blacktriangleright$} & $2.9 \times 10^{-11}$ & 1.0\\
      ethanol (95\%)  & $1.2$ & 789 & 22 &2.3 & \large{\color{ethanol}$\blacktriangleleft$}& $1.9 \times 10^{-9}$ &0.87\\
      & & & & & & & \\
      WG3 &3.0 &1100 &$67 \pm 3$ &2.3 &\color{eau_gluc_1}\large{$\bullet$}& $2.1 \times 10^{-9}$ &0.37\\
      WG13 &13 &1187 &$67 \pm 3$ &2.3 &\color{eau_gluc_2}\large{$\bullet$}& $6.9 \times 10^{-7}$ &0.07\\
      WG24 &24 &1208 &$67 \pm 3$ &2.3 &\color{eau_gluc_3}\large{$\bullet$}& $4.6 \times 10^{-6}$ &0.03\\
      & & & & & & & \\
      WU8  &8 &1011 & $53 \pm 1$ &2.3 &\color{eau_ucon_8}\large{$\vardiamondsuit$}& $3.2 \times 10^{-7}$ &0.10 \\
      WU17  &17 &1020 &$52 \pm 1$ &2.3 &\color{eau_ucon_17}\large{$\vardiamondsuit$}& $6.4 \times 10^{-6}$ &0.03\\
      WU42  &42 &1032 & $52 \pm 1$&2.3 &\color{eau_ucon_42}\large{$\vardiamondsuit$}& $2.1 \times 10^{-4}$ & 0.01\\
      WU80  &80 &1041 & $52 \pm 1$ &2.3 &\color{eau_ucon_80}\large{$\vardiamondsuit$}& $3.1 \times 10^{-3}$ & 4.9 $\times 10^{-3}$\\
      WU140  &140  &1048 &$51 \pm 1$ &2.3 &\color{eau_ucon_140}\large{$\vardiamondsuit$}& $2.9 \times 10^{-2}$ & 1.4 $\times 10^{-3}$\\
      WU152 &152 &1051 &$51 \pm 1$ &2.3 &\color{eau_ucon_152}\large{$\vardiamondsuit$}& $4.8 \times 10^{-2}$ & 1.1 $\times 10^{-3}$\\
      WU210  &210 &1057 &$50 \pm 1$ &2.3 &\color{eau_ucon_210}\large{$\vardiamondsuit$}& $1.7 \times 10^{-1}$ & 5.8 $\times 10^{-4}$\\
      WU260  &260 &1058 &$49 \pm 1$ &5.2 & \color{eau_ucon_260}\large{$\vardiamondsuit$} & $3.2 \times 10^{-1}$ & 9.0 $\times 10^{-3}$\\
      WU620  &620 &1066 &$47 \pm 1$ &2.3 &\color{eau_ucon_620}\large{$\vardiamondsuit$}& $1.3 \times 10^1$ & 9.6 $\times 10^{-5}$\\
      WU930  &930 &1074 &$47 \pm 1$ & 2.0 &\color{eau_ucon_930}\large{$\vardiamondsuit$}& $6.5 \times 10^1$ & 1.8 $\times 10^{-5}$\\
      WU1120  &1120 & 1075& $46 \pm 1$ &2.3 &\color{eau_ucon_1120}\large{$\vardiamondsuit$}& $1.1 \times 10^2$ & 3.0 $\times 10^{-5}$\\
      WU2890  & 2890& 1085 & $45 \pm 1$ &2.0 &\color{eau_ucon_2890}\large{$\vardiamondsuit$}& $7.4 \times 10^3$ & 1.7 $\times 10^{-6}$\\
      
      & & & & & & & \\
      WT2700  &2700 &1187 &$32 \pm 1$ &5.2 &\color{Triton}$\blacksquare$ & $1.6 \times 10^4$ & 1.3 $\times 10^{-4}$\\
  \end{tabular}
  \end{center}
    \caption{Liquid properties and cell gap $h$. WUxx, WGxx and WTxx stand for water mixtures with \textsc{Ucon}, glucose and Triton/ZnCl, respectively, where xx is the corresponding viscosity in mPa~s. The uncertainty is 1~kg.m$^{-3}$ for density, 0.1~mm for $h$ and for viscosity 10\% when $\eta < 20$~mPa~s and 1\% otherwise. $T$ is the time required to rise up to the surface for the fastest bubble.}
    \label{sol}
\end{table}

\section{Experimental set-up}\label{sec:experimental_set_up}

We investigate the rising motion of a single bubble in a Newtonian liquid initially at rest confined between two vertical plates (Hele-Shaw cell {$L_c =20$}~cm large, {$H_c=30$}~cm high, {see figure \ref{fig:chrono_v3}(a)}). Experiments are performed using different cell gaps ($h=[2.0, 2.3, 5.2]$~mm). The recent study of \cite{Pavlov_2021} on the role of lateral confinement on bubble motion shows that there is very little effect of the width-to-length ratio in our setup. A wide variety of liquids are considered with different viscosity, density and surface tension (Table~\ref{sol}). The viscosity is taken from handbooks for water and ethanol while for the other solutions, it is measured by a \textsc{Malvern Kinexus Ultra+} rheometer at shear rates close to the experimental conditions and at room temperature (from 20$^\circ$C to 25$^\circ$C). The surface tension and liquid density are quantified using pendant drop method in a \textsc{Attension Theta} tensiometer and a \textsc{Anton Paar DMA 35} densimeter, respectively. The Morton number $\Mor=g\eta^4/(\rho \gamma^3)$, which characterises the liquid physical properties, varies over 15 orders of magnitude.

Bubbles are generated at the centre of the cell's bottom with the help of a millimetric-sized pipe attached to a manually-controlled 50~mL syringe.
The cell is backlit uniformly with a LED panel while a computer-controlled {camera} (Basler AC-0.400, 2048 x 2048 pixels) records, for each run, the rising motion of the bubble at 10 to 60 fps (depending on the bubble velocity). {Note that although we are able to visualize the whole cell, the bubble motion is analyzed in a region of interest far from the cell boundaries ($\approx$ 4 cm from the top and bottom, and 8 cm from the sides as the bubble roughly rise vertically).} Images are then binarised {(the threshold of binarization induces an error of less than 1\% on the bubbles characteristics)} and standard techniques in \textsc{Matlab}$\circledR$ are performed to identify the bubble contour, define the equivalent ellipse and compute the bubble speed $v_b$ and aspect ratio $\chi=a/b$. We remind that $a$ and $b$ are the bubble semi-axes parallel (longitudinal) and perpendicular (transverse) to its motion, respectively. The equivalent planar bubble diameter is therefore $d_2=2\sqrt{ab}$. {In the following, only bubbles with $d_2>h$ are considered.}

\section{Experimental Results}\label{sec:experimental_results}
\subsection{General observations}
\begin{figure}[ht!]
      \includegraphics[width=0.9\linewidth]{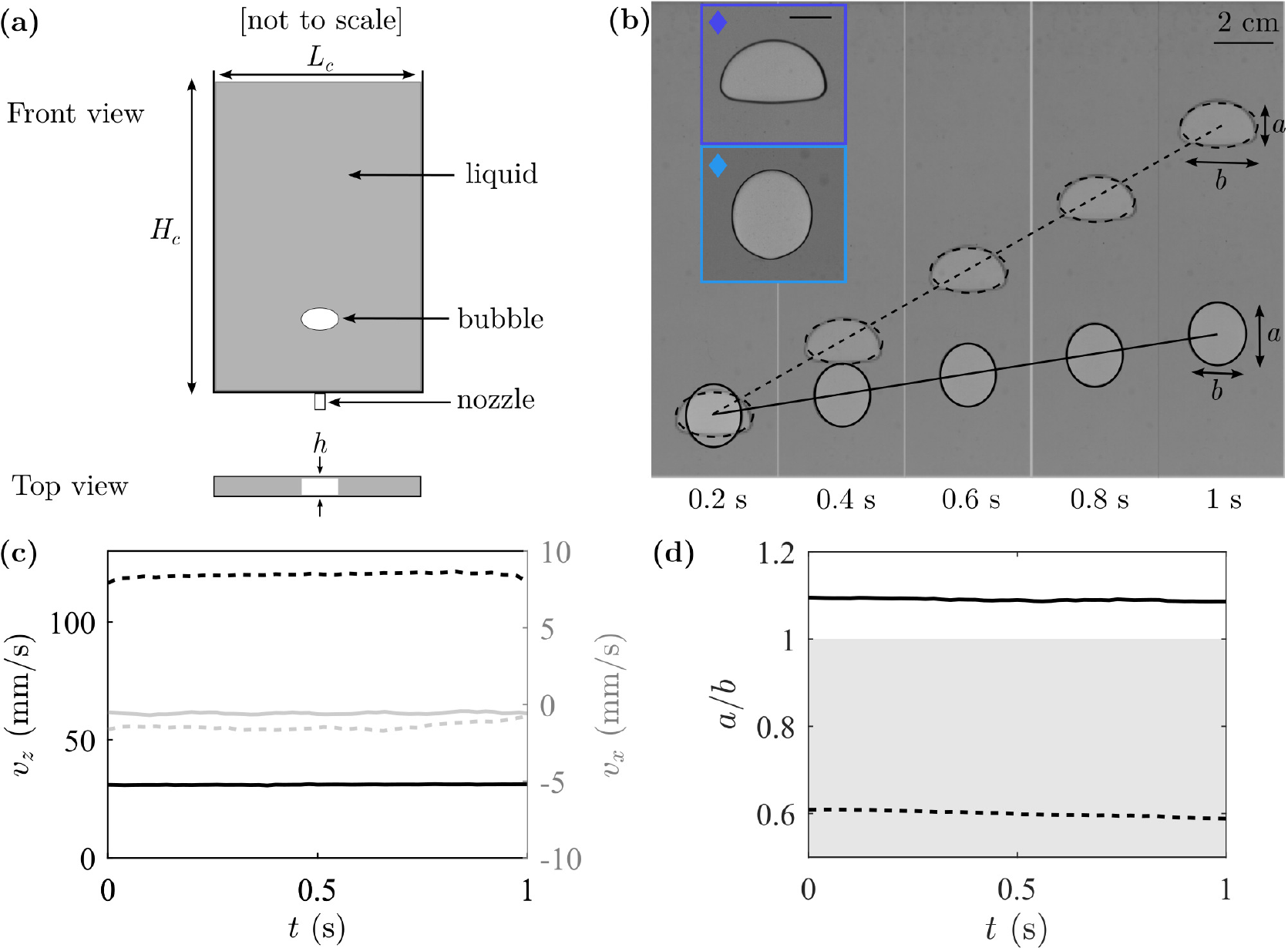}
  \caption{{\textbf{(a)} Schematic of the experimental setup.} \textbf{(b)} Chronophotography of two individual bubbles [$h=2.3$~mm] illustrating a slower, elongated bubble (WU140, solid line, $\mbox{Re}_{2h}=7.2 \times 10^{-2}$) in contrast to a faster, flattened bubble (WU17, dashed line, $\mbox{Re}_{2h}=2$). {Images here are cropped.} See supplemental material. Inset: zoom on each bubble indicating the scale (black line 1~cm). Temporal evolution of \textbf{(c)} the horizontal ($v_x$, grey) and vertical ($v_z$, black) velocity and \textbf{(d)} the aspect ratio $\chi=a/b$. The grey area corresponds to $\chi<1$.}
\label{fig:chrono_v3}
\end{figure}
 A chronophotograph of two rising bubbles of almost identical apparent diameter $d_2$ is displayed in figure~\ref{fig:chrono_v3}(b). The bubbles rise in Water/\textsc{Ucon} mixtures whose viscosity differs by an order of magnitude. As expected, the bubble in the more viscous liquid rises slower. The bubble shape is clearly different: an oval-shaped bubble elongated vertically is observed in the viscous liquid (WU140) while a oblate, flattened bubble is seen in the less viscous liquid (WU17). In the former case, the bubble shape resembles very much an ellipse differing only by a small cusp at the rear (see insert). In the latter case, the bubble does not display an ideal elliptic shape.
 
 The temporal evolution of the bubble velocity $v_b$ and its aspect ratio $\chi=a/b$ are given in figure~\ref{fig:chrono_v3}(c) and \ref{fig:chrono_v3}(d), respectively. The vertical velocity ($v_z$, black) is at least two orders of magnitude larger than the horizontal speed ($v_x$, grey) for both bubbles. No zigzag motion is reported. Neither $v_z(t)$, $v_x(t)$ nor the aspect ratio $\chi (t)$ vary significantly during the bubble rise. This is the case for most bubbles under study since the time scale $\tau = h^2\rho/(4\eta)$ required to establish a steady rising motion (\cite{Filella_2015}) is much smaller than the time required to rise up to the surface (the ratio between these two characteristic times is given in the last column of Table~\ref{sol}, where $T$ is the time required to rise up to the surface for the fastest bubble). The condition is however not attained in water and ethanol. Unless specified, all quantities, namely, the bubble speed, aspect ratio and diameter, provided in the following sections are time-averaged (${\langle \cdot \rangle}$) {that corresponds to a rise of 20 cm at most}. The errorbars indicate the standard deviation from these average quantities {and are present on every graph; they are often smaller than the symbols size}.
 
In summary, for given liquid mixtures that differ only by their viscosity, these observations clearly indicate that the bubbles are either elongated or flattened as the associated Hele-Shaw Reynolds number changes from $7 \times 10^{-2}$ to 2. We therefore further investigate the bubble speed in WU-mixtures by properly controlling the liquid viscosity so that $\mbox{Re}_{2h}$ varies while keeping the surface tension approximately constant.

\subsection{Bubble rising speed }
\label{subsec:Vertical_velocity}
\begin{figure}[ht!]
  \begin{center}
  \includegraphics[width=0.95\linewidth]{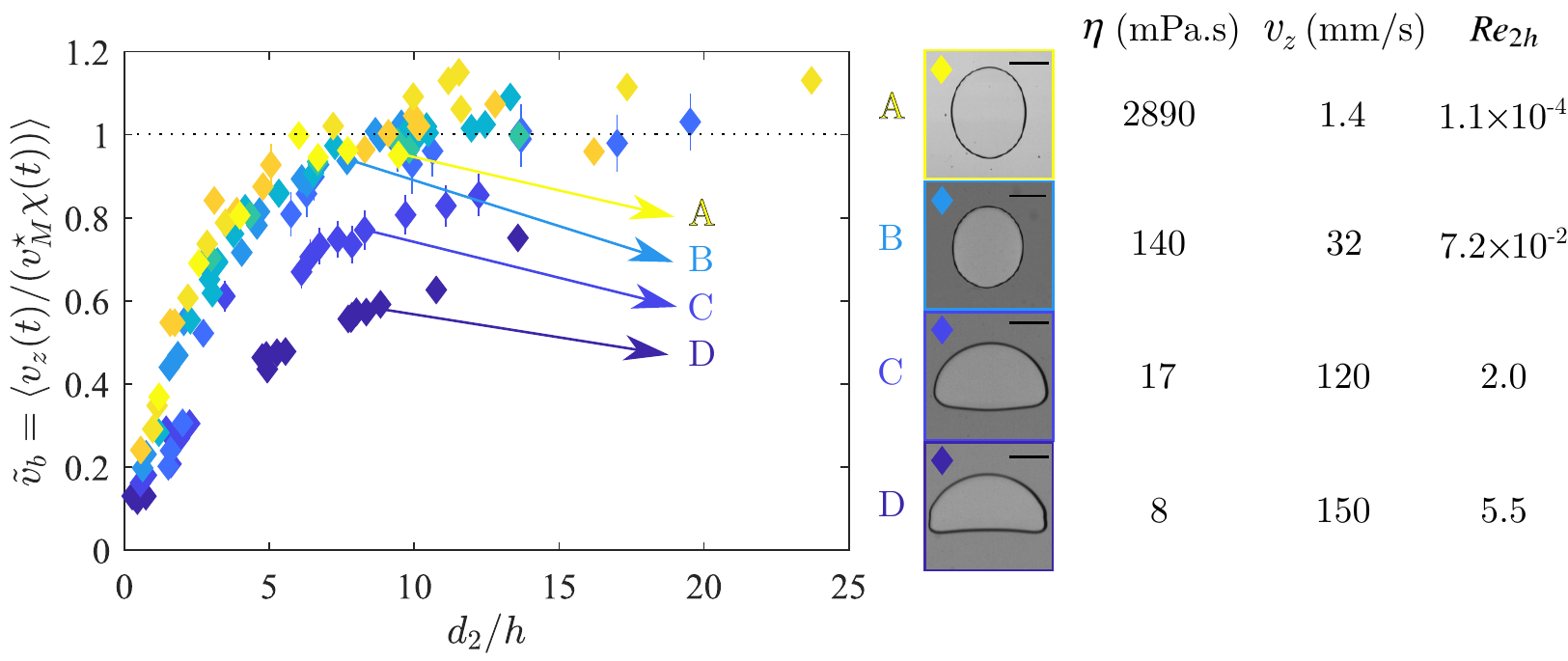}
  \end{center}
  \caption{Evolution of $\Tilde{v}_b=<v_z(t) / (v_M^{\star} \chi(t))>$ with $d_2/h$ [$h=2.0$ or 2.3~mm, see Table~\ref{sol}] for WU-mixtures with different liquid viscosity ({see right side of the figure and Table~\ref{sol}}). The dark line on the images corresponds to 1 cm. Images correspond to four typical bubbles of almost the same size along with their respective speed and Reynolds number.}
\label{fig:vbvm}
\end{figure}

As proposed by \cite{Taylor_1959} and \cite{Maxworthy_1986}, for $\mbox{Re}_{2h} \ll 1$, the time-averaged bubble speed should be given by equation~\ref{eq:velocity_maxworthy} when $d_2 \gg h$. We compute the time-averaged normalised bubble velocity $\tilde{v}_b = <v_z(t) / v_M(t)>$, where $v_M(t) = v_M^{\star} \chi(t)$ is a function of time as the bubble aspect ratio is free to evolve  during bubble rise (see figure~\ref{fig:chrono_v3}c). This quantity is displayed in figure~\ref{fig:vbvm} as a function of the time-averaged normalised bubble diameter, $d_2/h$.

For viscous water/\textsc{Ucon} mixtures ($\eta \gtrsim 100$~mPa.s), the normalised bubble speed $\tilde{v}_b$ increases monotonically from zero and plateaus at about unity at $d_2/h > 8$.  
For less viscous liquids, the trend is similar but the normalised bubble speed is smaller for comparable $d_2/h$. In addition, the plateau value of $\Tilde{v}_b$ is smaller than unity. Also in figure~\ref{fig:vbvm}, typical Reynolds numbers $\mbox{Re}_{2h}$ are given. A very good agreement between the experimental data and the theoretical viscous bubble speed (Eq.~\ref{eq:velocity_maxworthy}) is observed for large bubbles at $\mbox{Re}_{2h} \ll 1$ without any adjustable parameter. Equation~\ref{eq:velocity_maxworthy} even provides a reasonable estimation when $\mbox{Re}_{2h} \simeq 1$. {Nonetheless, as $Re_{2h}$ increases, inertia becomes important and the bubble speed deviates from the viscous limit, $v_M$. This is due to secondary flows around the bubble as $Re=(d_2/h)^2Re_{2h}=\rho v_b d_2/\eta$ becomes large \citep{Bush_1998, pavlov2021}}.

Typical bubble shapes at different Reynolds numbers $\mbox{Re}_{2h}$ are displayed for a fixed normalised diameter $d_2/h \simeq 10$. Note that bubbles B and C correspond to the examples shown in figure~\ref{fig:chrono_v3}. 
As the liquid viscosity is decreased while keeping the surface tension unchanged, the bubble shape continuously evolves from a longitudinally-elongated quasi-elliptic contour to a flattened oblate bubble. This indicates a decrease in the bubble aspect ratio as $\mbox{Re}_{2h}$ increases.
 
\begin{figure}[ht!]
%  \centerline{\includegraphics[width=0.8\linewidth]{fig3.eps}}
  \centerline{\includegraphics[width=0.8\linewidth]{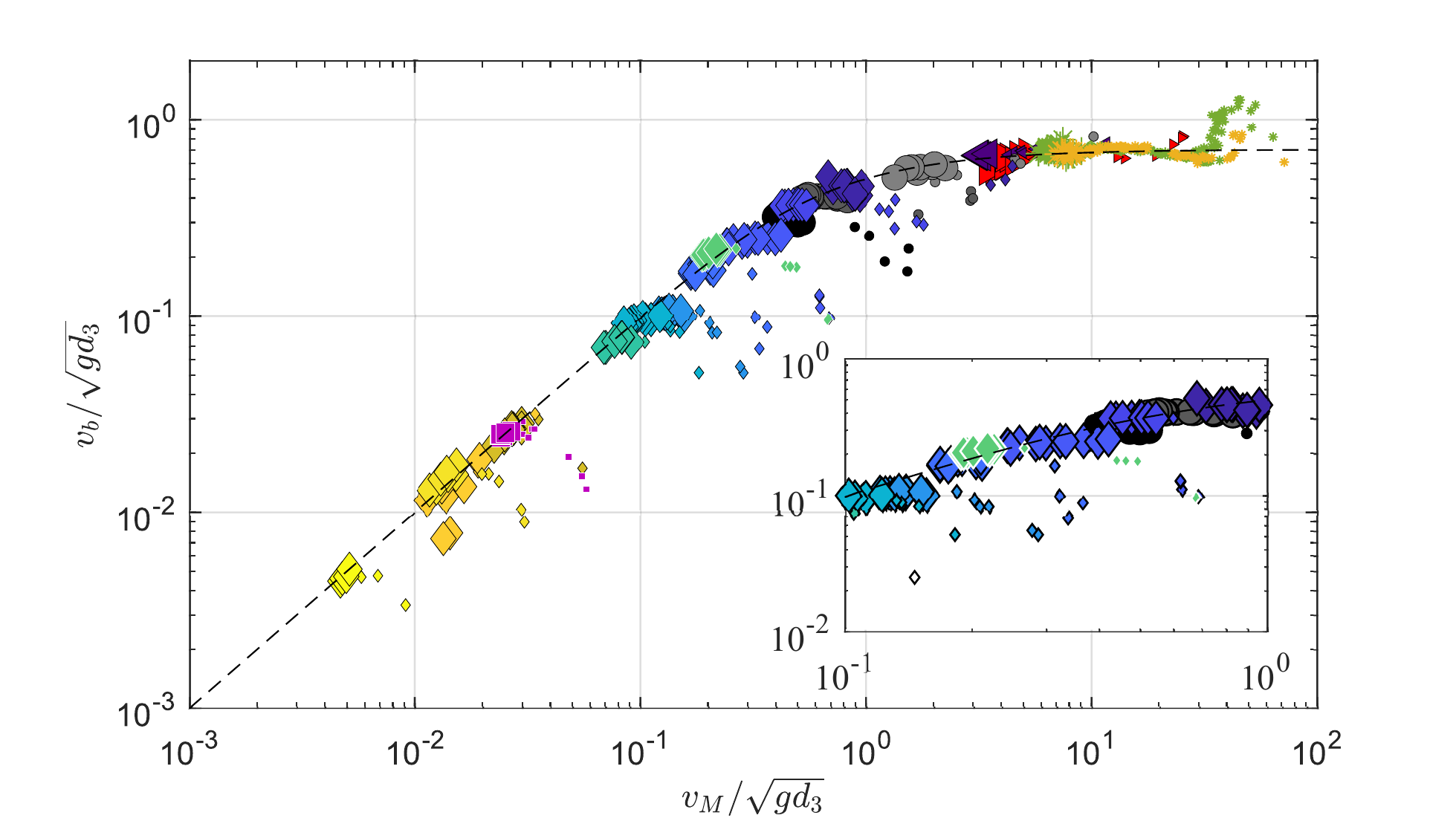}}
  \caption{Experimental velocity ratio $v_b/\sqrt{gd_3}$ as a function of $v_M/\sqrt{gd_3}$. Small (resp. large) symbols indicate bubbles with $d_2/h<4$ (resp $d_2/h>4$). Colors are the ones defined in Table~\ref{sol}. Data with dark and white edges correspond to $h=2.3~$mm and $h=5.2~$mm, respectively. Stars at high $v_M/\sqrt{gd_3}$ are data from~\cite{Filella_2015} (\textcolor{Filella}{$\ast$}) and \cite{Pavlov_2021} (\textcolor{Pavlov}{$\ast$}). Insert: zoom on the transition. The dashed line is the prediction from equation~\ref{eq:v_bv_M} with $\beta=3.9 \pm 0.1$.}
\label{fig:vbgd3}
\end{figure}

These results strongly suggest that inertia modifies not only the maximum bubble speed but also the bubble aspect ratio. An appropriate model to improve the theoretical result of \cite{Taylor_1959} (equation~\ref{eq:velocity_maxworthy}) should therefore include inertial effects. For large bubbles at $\mbox{Re}_{2h} \ll 1$ at dynamic equilibrium, \cite{Maxworthy_1986} derived this expression by balancing the injected power $P_b=\Delta \rho (\pi a b h) v_b$ with the viscous dissipation rate $\Dot{\Phi}_v=12 \eta v_b^2 \pi b^2/h$ due to the viscous flow generated by the rising bubble.
Building upon this analysis, we propose that in addition to the viscous dissipation, the injected power $P_b$ should also contribute to overcome inertial effects. The latter, in general, are proportional to the kinetic energy of an equivalent liquid volume that is set in motion by the bubble $1/2 \rho v_b^2(\pi a b h)$ and the characteristic time scale during which the bubble volume exchanges energy with the liquid, i.e, $d_3/v_b$ where $d_3$ is the volume-based diameter as in equation~\ref{eq:inertial_speed}. Thereby, the modified power balance leads to
\begin{equation}\label{eq:power_equilibrium}
    \rho g(\pi abh)v_b = \frac{12 \pi \eta v_b^2b^2}{h} +  \frac{1}{2} \rho v_b^2 (\pi abh) \left( \beta \frac{v_b}{d_3} \right),
\end{equation}
where $\beta$ is an arbitrary constant. Note that the ratio between the two terms on the right-hand side is $(\beta/24)\chi \Re_{3h}$, where $\mbox{Re}_{3h}=(\rho v_b d_3/\eta) (h/d_3)^2=\Re_{2h} (d_2/d_3)$ is a volume-based Hele-Shaw Reynolds number. For $\mbox{Re}_{3h} \ll 1$, the last term is negligible so equation~\ref{eq:power_equilibrium} gives $v_b=v_M$ whereas, for $\mbox{Re}_{3h} \gg 1$, the viscous dissipation can be ignored with respect to inertia-added power, which leads to $v_b=\sqrt{2/\beta}\sqrt{gd_3}$, regardless of the aspect ratio. Furthermore, rewriting~\ref{eq:power_equilibrium},
\begin{equation}
    v_b=\frac{2v_M}{1+\sqrt{1+2\beta \left(\frac{v_M}{\sqrt{gd_3}}\right)^2}} ,
    \label{eq:v_bv_M}
\end{equation}
where the ratio $v_M/\sqrt{gd_3}$ is the parameter that distinguishes the viscous and inertial regimes for bubbles in a Hele-Shaw cell. In equation~\ref{eq:v_bv_M},  the only parameter depending on the aspect ratio is $v_M=v_M^{\star} \chi$. In figure~\ref{fig:vbgd3}, this new expression is now compared with experimental data for all liquids given in Table~\ref{sol} and also, for data from \cite{Filella_2015} and \cite{Pavlov_2021} for a different gap ($h=3.1~$mm). 

The dashed line is the prediction from equation~\ref{eq:v_bv_M} with $\beta = 3.9 \pm 0.1$ (best fit on large bubbles). {When $v_M/\sqrt{gd_3} \gg 1$ and $\beta=3.9$, equation~\ref{eq:v_bv_M} leads to $v_b \simeq 0.72 \sqrt{gd_3}$, which corresponds within the errorbars to the inertial limit obtained by~\cite{Filella_2015} (equation~\ref{eq:inertial_speed}).} All data {for} the large bubbles ($d_2 > 4h$) are in very good agreement with equation~\ref{eq:v_bv_M}. Smaller bubbles fall below the dashed curve as long as $v_M/\sqrt{g d_3} < 10$, indicating that the equation~\ref{eq:v_bv_M} provides probably an upper boundary for the bubble speed at $\mbox{Re}_{3h}\ll 1$. This is probably due to the fact that the viscous dissipation for small bubbles is overestimated in equation~\ref{eq:power_equilibrium}. {Note that the expression for the bubble volume $V=A h$ is not exact since, especially for small bubbles, the rounding of the edges gives a lower volume (for a bubble with $d_2=2h$, this gives an error of 5 \% on $\sqrt{d_3}$). Also, $A=ab$ is not exact as bubbles are not perfectly elliptical but this estimation is really satisfactory (less than 2\% of error in the worst case).}

In conclusion, the rise speed of large bubbles ($d_2 \gg h$) in a Hele-Shaw cell is uniquely determined by the ratio $v_M/\sqrt{gd_3}$ for all Morton numbers in our study. This is all the more surprising that bubbles at very large Reynolds numbers exhibit shape and path oscillations, as already observed in previous works \citep{Filella_2015, Pavlov_2021}. { Finally, these results suggest that the dissipation in the film between the bubble and the walls does not significantly influence the bubble speed when $d_2 \gg h$. This is consistent with previous works \citep{Toupoint_2021, Keiser_2018} since in our case the gas/liquid viscosity ratio is very small.}

\subsection{Bubble aspect ratio}
\label{subsec:Aspect_ratio}
\begin{figure}[ht!]
%  \centerline{\includegraphics[width=0.85\linewidth]{fig4.eps}}
  \centerline{\includegraphics[width=0.85\linewidth]{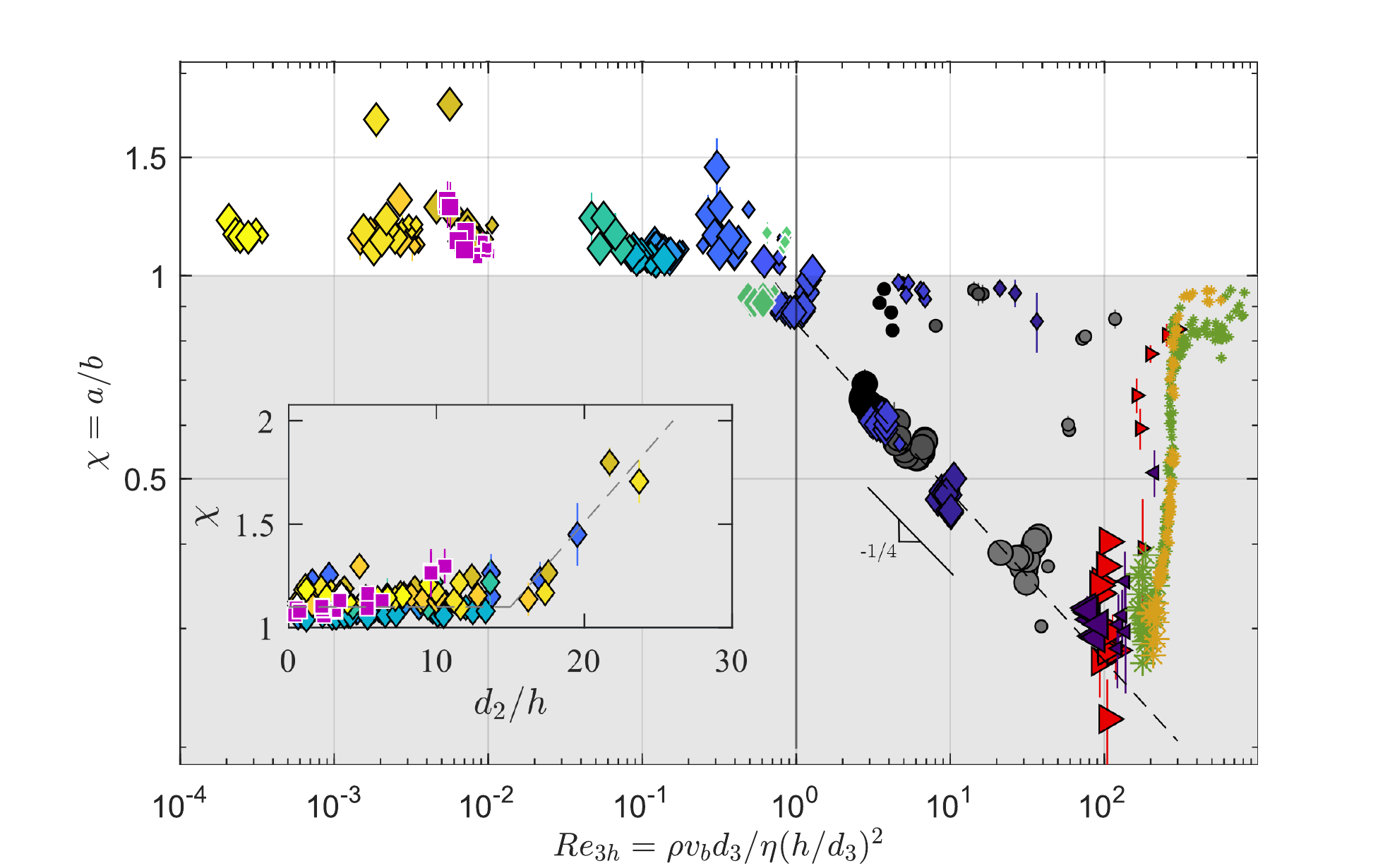}}
  \caption{Bubble aspect ratio $\chi$ as a function of $\mbox{Re}_{3h}$. Small (resp. large) symbols indicate bubbles with $d_2/h<4$ (resp $d_2/h>4$). Data with dark and white edges correspond to $h=2.3~$mm and $h=5.2~$mm, respectively. The grey area corresponds to $\chi<1$. The solid vertical grey line indicates $\mbox{Re}_{3h}=1$. Stars are data from~\cite{Filella_2015} (\textcolor{Filella}{$\ast$}) and \cite{Pavlov_2021} (\textcolor{Pavlov}{$\ast$}). Inset: $\chi$ as a function of $d_2/h$. The dashed line is a guideline. {The solid black segment with -1/4 indicates the slope of the dashed line.}}
\label{fig:RA}
\end{figure}
Figure~\ref{fig:RA} displays the time-averaged bubble aspect ratio $\chi=a/b$ as a function of $\mbox{Re}_{3h}$. 
All bubbles in the viscous regime ($\mbox{Re}_{3h} < 1$) are elongated in the direction of the bubble motion ($\chi \geq 1$). In contrast, bubbles in the inertial regime are flattened ($\chi <1$).

First, for Reynolds numbers $\mbox{Re}_{3h}$ above unity, the aspect ratio of large bubbles seems to decrease as $\mbox{Re}_{3h}^{-1/4}$. As the Reynolds number is further increased ($\mbox{Re}_{3h} \gtrsim {100}$), significant variations are reported, in particular for bubbles in ethanol (\textcolor{ethanol}{$\blacktriangleleft$}) and water (\textcolor{eau}{$\blacktriangleright$}). Interestingly, no such deviations occur for the corresponding bubble velocity (figure~\ref{fig:vbgd3}) since at sufficiently large velocity ratio ($v_M/\sqrt{g d_3} > 10$), the bubble speed depends only on its sphere-equivalent diameter $d_3$, irrespective of the aspect ratio. \cite{Filella_2015} accounted for the bubble aspect ratio variations by using the Weber number $\Web=\rho v_b^2 d_2/\gamma$, so that $\chi \approx \Web^{-1/2}$ for $1 < \Web < 10$. In our experiments, this scaling holds for water only but failed for all other liquids (not shown here).

Second, figure~\ref{fig:RA} (inset) presents the data corresponding to the viscosity-dominated regime ($\mbox{Re}_{3h} < 1$). As long as $d_2/h < 15$, the aspect ratio $\chi \simeq 1$, while for larger $d_2/h$, it increases with the normalised bubble diameter $d_2/h$. Finally, since the surface tension of WU and WT mixtures are not very different, it is not conclusive if this trend in aspect ratio for the low Reynolds number regime is universal. {Indeed, the scaling of the aspect ratio for large bubbles seems not to depend on the surface tension and thus on any dimensionless number involving it, such as the Bond number, the Capillary number or the Weber number}.

\subsection{Drag coefficient}

\begin{figure}[ht!]
%  \centerline{\includegraphics[width=0.75\linewidth]{fig5.eps}}
  \centerline{\includegraphics[width=0.75\linewidth]{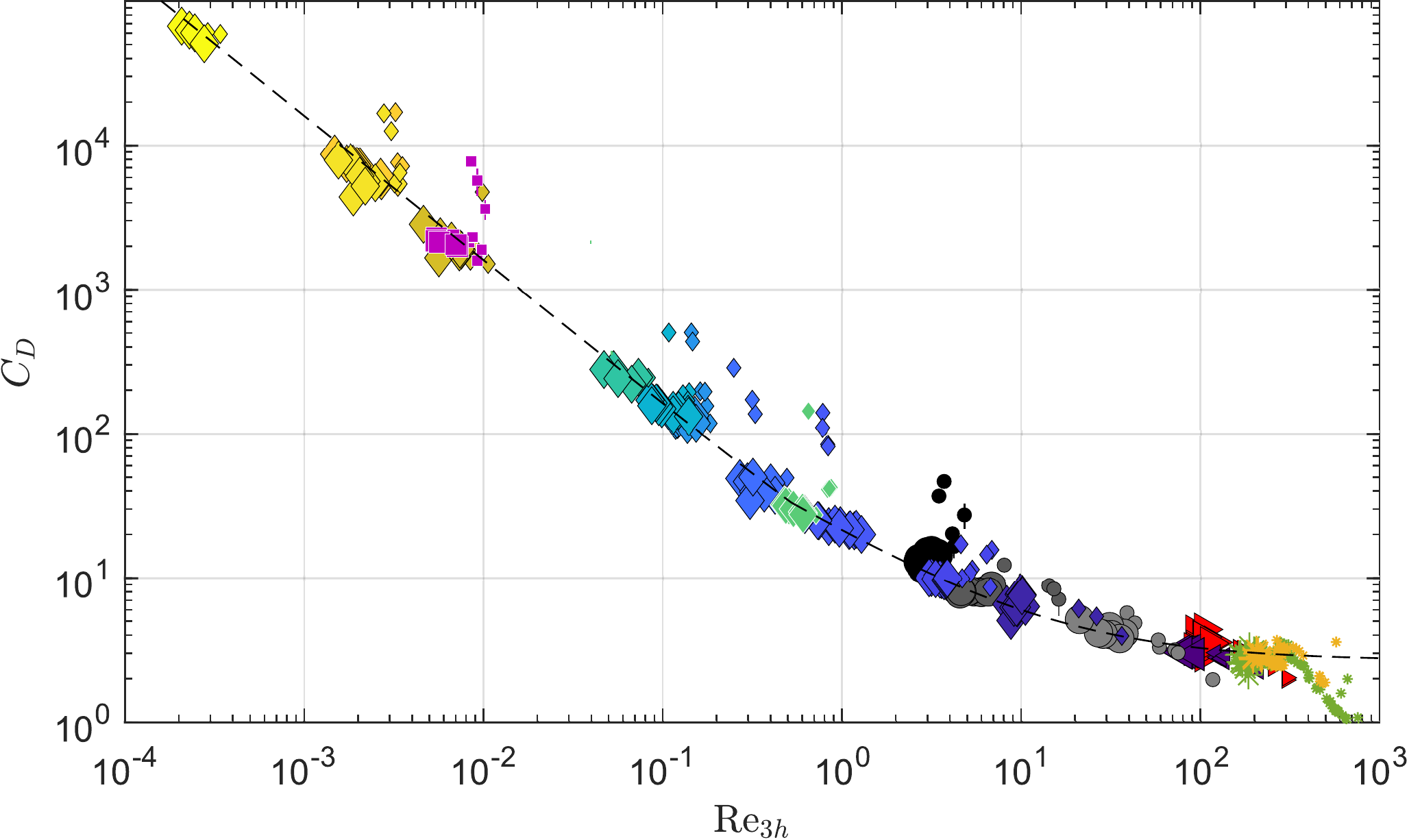}}
  \caption{Drag coefficient $C_{D}$ as a function of $\mbox{Re}_{3h}$. Small (resp. large) symbols indicate bubbles with $d_2/h<4$ (resp $d_2/h>4$). Colors are the ones defined in Table~\ref{sol}. Data with dark and white edges correspond to $h=2.3~$mm and $h=5.2~$mm, respectively. The dashed line corresponds to equation~\ref{eq:drag_coeff}. Stars at high $\mbox{Re}_{3h}$ corresponds to data from~\cite{Filella_2015} (\textcolor{Filella}{$\ast$}) and \cite{Pavlov_2021} (\textcolor{Pavlov}{$\ast$}).}
\label{fig:CD}
\end{figure}

From a dynamical point of view, the bubble's motion is characterised by its drag force $F_D$. It can be computed from the drag coefficient $C_D= F_D/(1/2 \rho v_b^2 S)$, where $S=\pi d_3^2/4$ the equivalent spherical surface and not the true projected area $2bh$ as often considered \citep{Filella_2015, Hashida_2019, Hashida_2020}.
At dynamic equilibrium, the drag force $F_D$ equals the driving force due to buoyancy $F_B = \rho g (\pi a b h)$ and so, the model developed in equation~\ref{eq:power_equilibrium} gives
\begin{equation}\label{eq:C3D}
    C_D=\frac{2\beta}{3} + \frac{16}{\Re_{3h}} \frac{1}{\chi},
\end{equation}
where the second term resembles the expression of the drag coefficient for an isolated spherical bubble in three dimensions, i.e $24/\Re_3$, where $\mbox{Re}_3=\rho v_b d_3/\eta$. However, for a large isolated bubble confined between two plates, $C_D$ is not only inversely proportional to a sphere-equivalent Hele-Shaw Reynolds number $\mbox{Re}_{3h}$ but also to its aspect ratio $\chi=a/b$. {Unlike the classical expression for the drag coefficient, it is necessary to either measure or model the bubble aspect ratio $\chi$ to estimate $C_D$ in equation~\ref{eq:C3D}. Since a general expression of $\chi$ is beyond the scope of our work,} we admit as inferred from section~\ref{subsec:Aspect_ratio} that $\chi \approx 1$ for $\mbox{Re}_{3h} < 1$ and $\chi=0.85 \Re_{3h}^{-1/4}$ for $\mbox{Re}_{3h} > 1$, {so that} the drag coefficient {becomes}
\begin{equation}\label{eq:drag_coeff}
  C_D = \left\{
    \begin{array}{ll}
      \dfrac{2\beta}{3} + \dfrac{16}{\Re_{3h}} & \Re_{3h} <1, \\[2pt]
      \dfrac{2\beta}{3} + \dfrac{16}{0.85\Re_{3h}^{3/4}}         & \Re_{3h} >1,
    \end{array} \right.
\end{equation}
with $\beta=3.9 \pm 0.1$ (see section~\ref{subsec:Vertical_velocity}). We now compare this expression (dashed line) with the experimental data (see figure~\ref{fig:CD}). Once again, for sufficiently large bubbles, all data collapse on the model, regardless of the cell gap $h$ and the liquid surface tension $\gamma$.

\section{Concluding remarks}
In this paper, we investigated the time-averaged bubble speed and bubble aspect ratio for large single bubbles rising in a vertical Hele-Shaw cell by properly controlling the bubble Reynolds number ($10^{-4} < \Re_{3h} < 300$) and the liquid Morton number $10^{-11} < \Mor < 10^4$.
For sufficiently large bubbles, we extended the classical power balance argument of \cite{Maxworthy_1986} by accounting for inertial effects to deduce $v_b=2v_M/(1+\sqrt{1+2 \beta (v_M/\sqrt{gd_3})^2})$, where $v_M=(\Delta \rho g h^2/(12\eta)) \chi$ with $\chi$ the bubble aspect ratio, $d_3$ the volume-based bubble diameter. The model fits well the experimental data with $\beta=3.9 \pm 0.1$. When $v_M/\sqrt{gd_3} \ll 1$, the bubble speed is given by the viscous bubble speed limit $v_b=v_M$ and conversely when $v_M/\sqrt{gd_3} \gg 1$, it tends towards the inertial limit $v_b=0.7\sqrt{gd_3}$, as already inferred by~\cite{Filella_2015, Hashida_2019}. The former corresponds to the viscous regime ($\mbox{Re}_{3h} \ll 1$) and the latter to the inertial regime ($\mbox{Re}_{3h} \gg 1$) wherein the bubble speed is proportional to $V^{1/6}$ \citep{Davies_1950, Collins_1965}. Unlike in 3D \citep{Maxworthy_1996}, the transition between these two limits is independent of the Morton number in the range given above. Our experimental data comprising a wide variety of liquids and cell gaps, along with data from previous studies, agree very well with our model as long as $d_2/h > 4$. 

In the viscous regime and also during the transition to the inertial regime, the aspect ratio is a necessary ingredient to correctly predict the bubble speed. In addition, at low Reynolds number ($\mbox{Re}_{3h} \lesssim 1$), only longitudinally elongated bubbles are reported in our experiments using water/\textsc{Ucon} and water/Triton solutions. Here, the bubble aspect ratio $\chi \approx 1$ for $d_2<15h$ and then linearly increases with $d_2/h$. No dependence on the liquid surface tension was observed but more experiments with other liquids should provide a conclusive answer. On the contrary, in the inertial regime, we reported flattened bubbles for all liquids, including water/\textsc{Ucon} mixtures such that $\chi \approx 0.85 \Re_{3h}^{-1/4}$. These results strongly suggest that the bubble {aspect ratio} is Reynolds number dependent {and it could be the signature of liquid inertia as it flows past the bubble \citep{Bush_1998, Filella_2015}}. Also, when the Reynolds number is sufficiently large, the time-averaged bubble aspect ratio shows strong deviations from the above scaling. 
This is probably related to the unsteady flow in the bubble wake along with surface tension and confinement effects, which are left for future investigations. {Also, the influence of surface tension on the bubble speed in the viscous regime needs to be further understood for smaller bubbles ($d_2<4h$)}

We acknowledge M. Moulin for experimental help. We thank L. Pavlov and P. Ern for sharing their experimental data and J.-P. Matas for useful discussions. This work was supported by the LABEX iMUST of the University of Lyon (ANR-10-LABX-0064), created within the program ``Investissements d'Avenir'' set up by the french government and managed by the French National Research Agency (ANR).

Declaration of Interests. The authors report no conflict of interest.

\bibliographystyle{plainnat}
\bibliography{SingleBubbleHeleShaw_ViscousToInertialRegimes_arXiv}

\end{document}